\documentclass[twocolumn]{aastex61}

\usepackage{amsmath}
\usepackage{units}
\usepackage{hyperref}

\newcommand{\MESA}{\texttt{MESA}}
\newcommand{\FUNS}{\texttt{FUNS}}

\newcommand{\gcc}{\ensuremath{\mathrm{g\,cm^{-3}}}} 
\newcommand{\Tc}{\ensuremath{T_{\mathrm{c}}}} 
\newcommand{\rhoc}{\ensuremath{\rho_{\mathrm{c}}}} 
\newcommand{\Ye}{\ensuremath{Y_{\mathrm{e}}}} 
\newcommand{\mpr}{\ensuremath{m_{\mathrm{p}}}} 
\newcommand{\Msun}{\ensuremath{\mathrm{M}_{\sun}}} 
\newcommand{\Msunyr}{\ensuremath{\rm \Msun\,yr^{-1}}} 

\newcommand{\MRPaper}{MR16}
\newcommand{\PPaper}{P17}

\newcommand{\nuclei}[2]{\ensuremath{\mathrm{^{#1}#2}}}

\newcommand{\carbon}[1][12]{\nuclei{#1}{C}}
\newcommand{\nitrogen}[1][14]{\nuclei{#1}{N}}
\newcommand{\oxygen}[1][16]{\nuclei{#1}{O}}
\newcommand{\fluorine}[1][19]{\nuclei{#1}{F}}
\newcommand{\neon}[1][20]{\nuclei{#1}{Ne}}
\newcommand{\sodium}[1][23]{\nuclei{#1}{Na}}
\newcommand{\magnesium}[1][24]{\nuclei{#1}{Mg}}
\newcommand{\aluminum}[1][27]{\nuclei{#1}{Al}}
\newcommand{\silicon}[1][28]{\nuclei{#1}{Si}}
\newcommand{\phosphorus}[1][31]{\nuclei{#1}{P}}
\newcommand{\sulfur}[1][32]{\nuclei{#1}{S}}
\newcommand{\chlorine}[1][35]{\nuclei{#1}{Cl}}

\newcommand{\potassium}[1][39]{\nuclei{#1}{K}}

\newcommand{\chromium}[1][52]{\nuclei{#1}{Cr}}
\newcommand{\manganese}[1][55]{\nuclei{#1}{Mn}}
\newcommand{\iron}[1][56]{\nuclei{#1}{Fe}}

\received{September 25, 2017}
\revised{November 10, 2017}
\accepted{November 10, 2017}
\submitjournal{ApJ}

\shorttitle{Carbon simmering}
\shortauthors{Schwab et al.}

\begin{document}


\title{Exploring the Carbon Simmering Phase: Reaction Rates, Mixing, and the Convective Urca Process}

\author[0000-0002-4870-8855]{Josiah Schwab}
\altaffiliation{Hubble Fellow}
\affiliation{Department of Astronomy and Astrophysics, University of California, Santa Cruz, CA 95064, USA}
\correspondingauthor{Josiah Schwab}
\email{jwschwab@ucsc.edu}

\author[0000-0002-1919-228X]{H\'{e}ctor Mart\'{i}nez-Rodr\'{i}guez}
\affiliation{Department of Physics and Astronomy and Pittsburgh Particle Physics, Astrophysics and Cosmology Center (PITT PACC), University of Pittsburgh, 3941 O'Hara Street, Pittsburgh, PA 15260, USA}

\author[0000-0001-6806-0673]{Anthony L. Piro}
\affiliation{The Observatories of the Carnegie Institution for Science, 813 Santa Barbara Street, Pasadena, CA 91101, USA}

\author[0000-0003-3494-343X]{Carles Badenes}
\affiliation{Department of Physics and Astronomy and Pittsburgh Particle Physics, Astrophysics and Cosmology Center (PITT PACC), University of Pittsburgh, 3941 O'Hara Street, Pittsburgh, PA 15260, USA}
\affiliation{Institut de Ci\`encies del Cosmos (ICCUB), Universitat de Barcelona (IEEC-UB), Mart\'i Franqu\'es 1, E08028 Barcelona, Spain}

\begin{abstract}
  The neutron excess at the time of explosion provides a powerful
  discriminant among models of Type Ia supernovae.  Recent
  calculations of the carbon simmering phase in single degenerate
  progenitors have disagreed about the final neutron excess.  We find that
  the treatment of mixing in convection zones likely contributes to
  the difference.  We demonstrate that in \MESA\ models, heating from
  exothermic weak reactions plays a significant role in raising the
  temperature of the WD.  This emphasizes the important role that the
  convective Urca process plays during simmering.  We briefly
  summarize the shortcomings of current models during this phase.
  Ultimately, we do not pinpoint the difference between the results
  reported in the literature, but show that
  the results are consistent with different net energetics of the
  convective Urca process.  This problem serves as an important
  motivation for the development of models of the convective Urca
  process suitable for incorporation into stellar evolution codes.

\end{abstract}


\section{Introduction}
\label{sec:intro}

Type Ia supernovae (SNe Ia) are the thermonuclear explosions of white 
dwarf (WD) stars destabilized by mass accretion from a close binary
companion. Despite their relevance for many 
fields of astrophysics, such as galactic chemical evolution \citep{Ko06, An16}, 
studies of dark energy \citep{Ri98,Pe99} and constraints on $\Lambda$CDM 
parameters \citep{Be14,Re14}, basic aspects of SNe Ia remain unclear, 
including the precise identity of their stellar progenitors and the 
mechanism that triggers the thermonuclear runaway. There are two main proposed
progenitor channels for SNe Ia: the single degenerate, where the WD companion is
 a non-degenerate star \citep[e.g.,][]{No84,Ha96,Han04}, and the double degenerate, where the WD companion
 is another WD \citep[e.g.,][]{Ib84,Si10,Ku13,Shen2017b}

In the single degenerate scenario for SNe Ia, a massive,
accreting WD approaches the Chandrasekhar mass 
($M_{\rm{Ch}} \, {\simeq} \, 1.4 \, M_{\odot}$).  As the density and
temperature of the core increase, it reaches central conditions at
which carbon burning to begins to occur.  This energy release leads to
the formation of a central convective zone and a carbon ``simmering''
phase which lasts for thousands of years \citep[e.g.,][]{Piro2008b}.  As the temperature
continues to increase, the burning becomes dynamical; this results in
the birth of a deflagration and subsequently the explosion of the WD
\citep[e.g.,][]{Woosley2004a, Malone2014a}.

The composition of the material at the time of explosion, specifically
 its electron fraction \Ye, influences the nucleosynthesis
of the explosion, in particular the production of neutron-rich
isotopes.  Measurements of the abundances of both intermediate-mass and 
iron-peak elements in supernova remnants and
the comparison with models \citep{Ba08,Park13,Ya15,MartinezRodriguez2017} has
the potential to provide information about the progenitors.
Understanding the amount of neutronization expected during the
different phases of the pre-explosion evolution is important
because it is an observational probe sensitive to
the mass of the exploding WD and to the presence or absence
of an extended accretion phase leading to the thermonuclear runaway.

\citet{Piro2008a} and \citet{Chamulak2008} discuss the important
reactions involved in the simmering phase. By way of analytic
calculations and one-zone burns, they give estimates of the amount of
neutronization expected during simmering.
\citet{MartinezRodriguez2016}, hereafter \MRPaper, used \MESA\ to
perform simulations of accreting WDs through the carbon simmering
phase, for a range of metallicities, accretion rates, and cooling
ages.  One of the key findings of this study was that the
neutronization during simmering was less than the estimates of
\citet{Piro2008a} and \citet{Chamulak2008}.  This result was
understood as consequence of using a full stellar model: the central
convection zone spans several density scale heights, meaning that the
net electron capture rate is lower than one would estimate using the
central density.

Recently, \citet{Piersanti2017}, hereafter \PPaper, also evolved full
stellar models of accreting WDs through the simmering phase using the \FUNS\
code \citep{Straniero2006, Cristallo2009}.  This
work found values for the neutronization during simmering that were a
factor of $\approx 5$ greater than \MRPaper\ (at solar metallicity).
In addition, they found a significant dependence of the neutronization
on the metallicity of the WD, unlike \MRPaper\ and in contrast to the
proposal of \citet{Piro2008a} that simmering imparts a
metallicity-independent floor to the neutronization.  \PPaper\ did not
undertake a detailed comparison with other models in the literature
and were unable to clearly identify the key inputs which led their
models to differ from previous work.

In this paper, we work to identify the most important aspects of the
initial conditions and the input physics that differ between the
\MRPaper\ and \PPaper\ models.  In Section~\ref{sec:simmering} we
discuss the role of the nuclear reaction rates.  In
Section~\ref{sec:mesa-mixing}, we discuss the role of the mixing
algorithms.  We demonstrate that neither of these can explain the
difference between \MRPaper\ and \PPaper.  In
Section~\ref{sec:end-simmering}, we discuss the role of the convective
Urca process and the way in which its effects are included in stellar
evolution codes.  We show that the differences between \MRPaper\ and
\PPaper\ are could arise due to differences in the net effect of the
convective Urca process.  In Section~\ref{sec:conclusion} we conclude.
The files necessary to reproduce our results will be made publicly
available on \url{http://mesastar.org}.

\section{Effects of differing input physics}

In the conclusion of P17, the authors suggest a number of areas where
there are (or may be) differences in the input physics between their
models and the models in the literature.  In this section, we explore
the effects of some of these differences by running models similar to
those in MR16, but varying one aspect of the input physics at a time.

\subsection{Reaction rates and initial composition}
\label{sec:simmering}

\PPaper\ used a larger nuclear network than \MRPaper.  In particular,
\PPaper\ included additional weak reactions, beyond the
\sodium[23]-\neon[23] and \magnesium[25]-\sodium[25] Urca pairs considered in
\MRPaper.  However, Figure 1 and Table 2 of \PPaper\ demonstrates that
the \neon[21]-\fluorine[21], \sulfur[32]-\phosphorus[32]-\silicon[32],
and \iron[56]-\manganese[56]-\chromium[56] weak reactions that were
neglected in \MRPaper\ are of only moderate importance and cannot by
themselves explain the differences between the two sets of results.
Other isotopes neglected by \MRPaper\ (i.e. \fluorine[19],
\phosphorus[31], \chlorine[37], \potassium[39]) have mass fractions
$\lesssim 10^{-6}$ and so cannot be responsible for the differences
either.\footnote{For previous work discussing the effects of many Urca pairs,
  see e.g., \citet{Tsuruta1970}, \citet{Paczynski1973a},
  \citet{Iben1978a} and Chapter~11 in \citet{Arnett1996}.}

In order to shed light on the source of the discrepancy, we construct
simple \MESA\ models.  We construct homogenous WDs using the
composition listed in Table~\ref{table:toy-abundances}.  This
composition uses the abundances given in \PPaper\ for the most
important odd mass number isotopes and the central abundances from the
0.85 \Msun\ WD model used in \MRPaper\ for the even mass number isotopes.  (We choose the 0.85 \Msun\ model since the initial WD model used in \PPaper\ is a 0.82 \Msun\ WD.)
We then allow these models to accrete at $\unit[10^{-7}]{\Msunyr}$ and
evolve through the carbon simmering phase, up until
$\Tc = \unit[8\times10^8]{K}$.

\begin{deluxetable*}{lllllllll}
  \centering
  \caption{Abundances in our homogenous WD models.\label{table:toy-abundances}}

  \tablehead{
  \colhead{\carbon[12]} &
  \colhead{\oxygen[16]} &
  \colhead{\neon[20]} &
  \colhead{\neon[21]} &
  \colhead{\neon[22]} &
  \colhead{\sodium[23]} &
  \colhead{\magnesium[24]} &
  \colhead{\magnesium[25]} &
  \colhead{\aluminum[27]}
}
\startdata
4.05$\times 10^{-1}$ & 5.76$\times 10^{-1}$ & 1.34$\times 10^{-3}$ & 3.74$\times 10^{-5}$ & 1.37$\times 10^{-2}$ & 1.42$\times 10^{-4}$ & 4.42$\times 10^{-5}$ & 3.84$\times 10^{-5}$ & 5.60$\times 10^{-5}$ \\
\enddata
\tablecomments{ Odd mass
number abundances are from the ZSUN model in \PPaper. Even mass number
abundances are from the center of the 0.85 \Msun WD model in
\MRPaper, except for \oxygen[16], which is adjusted to ensure the abundances sum to 1.}
\end{deluxetable*}

\begin{figure}
  \centering
  \includegraphics[width=\columnwidth]{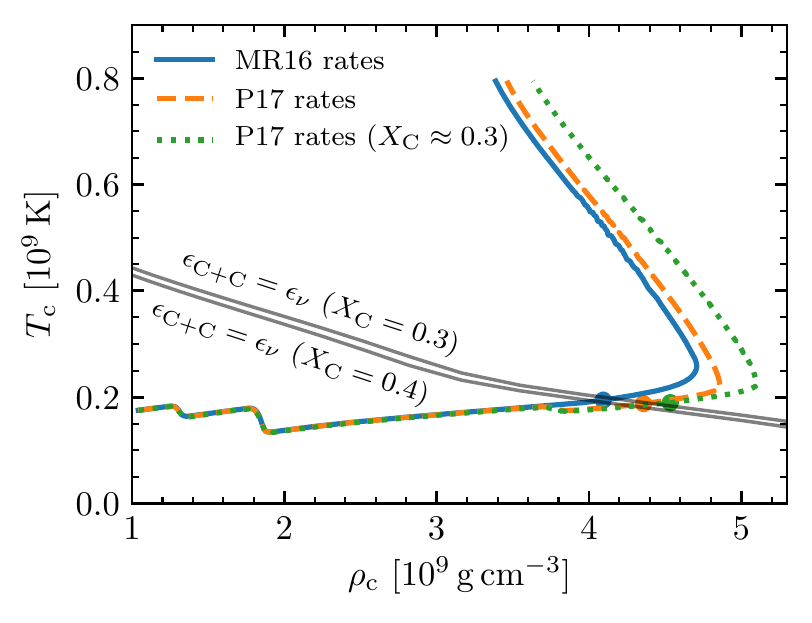}
  \caption{Central temperature vs.~density for models using different
    reaction rates.  Circles mark the beginning of simmering. The only
    significant difference is the effect of the
    \neon[21]-\fluorine[21] Urca pair at
    $\rho_{\rm c}\approx \unit[3.8\times10^9]{\gcc}$, which was
    neglected by \MRPaper.  The labeled grey lines show carbon
    ignition curves; above these lines, the energy release from carbon
    fusion exceeds thermal neutrino losses.}
  \label{fig:rates}
\end{figure}

One difference between \MRPaper\ and \PPaper\ is the adopted nuclear
reaction rates.  \MRPaper\ used the ``on-the-fly'' weak reaction
capabilities of \MESA\ as described and validated in
\citet{Paxton2015, Paxton2016}.  These capabilities calculate the weak
reaction rates without recourse to tabulations by using nuclear data
drawn from the literature and numerically evaluating the phase space
integrals at runtime each time a rate is needed. This is appropriate
when only a few transitions dominate the rate, as is the case here.
It circumvents difficulties associated with interpolation in tables
that have large changes in the rate between adjacent points
\citep{Fuller1985, Toki2013}.  \PPaper\ used the tabulations of
\citet{Suzuki2016} and a table for the
$\nitrogen[13](e^-, \nu_e)\carbon[13]$ reaction constructed using the
measurements from \citet{Zegers2008}.  These tables use a fine grid of
densities and temperatures so that they do not suffer from the
aforementioned interpolation issues.

In order to test differences associated with these varying sources of
the reaction rates, we incorporated the \citet{Suzuki2016} tables and
the $\nitrogen[13](e^-, \nu_e)\carbon[13]$ table used in \PPaper\
(G. Mart\'{i}nez-Pinedo, private communication) into \MESA.  In
Figure~\ref{fig:rates} we show the results of \MESA\ models with the
\MRPaper\ rates (solid line) and the \PPaper\ rates (dashed line).
The only significant difference between the two models is the effect of
the \neon[21]-\fluorine[21] Urca pair (neglected by \MRPaper), which
provides enough cooling to create a small shift in the ignition
density.  This leads to a $\approx 10 \%$ change in the central
neutron excess.  This demonstrates that these rates are not a
significant contributor to the differences between these results.

Carbon ignition occurs when the center of the WD reaches conditions
where the energy release from carbon burning exceeds the thermal
neutrino losses.  Such a condition can be defined in terms of the
neutrino loss rates, the carbon mass fraction, and the
\carbon[12]-\carbon[12] reaction rate.  \MRPaper\ and \PPaper\ use the
carbon burning rate from \citet{Caughlan1988}.  Under these
conditions, the reaction rate is non-negligibly influenced by the
screening treatment \citep[e.g.,][]{Yakovlev2006}.  The screening
treatment in \FUNS\ is described in \citet{Chieffi1998}: for the weak,
intermediate and intermediate-strong regimes it uses the electron
screening provided by \citet{Dewitt1973} and \citet{Graboske1973}
while for the strong regime it uses electron screening provided by
\citet{Itoh1977} and \citet{Itoh1979}.  The screening treatment in
\MESA\ is described in \citet{Paxton2011}: it combines
\citet{Graboske1973} in the weak regime and \citet{Alastuey1978} with
plasma parameters from \citet{Itoh1979} in the strong regime.
Nominally then, the carbon burning rate should be quite similar.
However, carbon ignition occurs at higher density in the \PPaper\
models, at $\rhoc \approx \unit[5\times10^9]{\gcc}$ in their ZSUN
model.

The higher ignition density for the \PPaper\ models is in part due to
the lower central carbon mass fraction in those models.  The \PPaper\
ZSUN model has $X_{\rm C} = 0.29$ (L.~Piersanti, private
communication).  This shifts the ignition density appreciably.
Figure~\ref{fig:rates} shows ignition curves for these values of the
carbon fraction and a model with the fiducial composition, except the
C/O ratio has been changed to give $X_{\rm C} = 0.3$.  This model has
a central neutron excess 10\% lower than the model with the MR16 rates
and the higher carbon fraction.  This difference is in the wrong
direction to account for the discrepancy between the results of \MRPaper\ and \PPaper.
We note that the difference in central carbon fraction does not
completely explain the difference in ignition densities:
the \PPaper\ ZSUN model ignites at a $\approx 10\%$ higher density
than our analogous model.  Thus there is likely some difference in the
input physics relevant to carbon ignition between the two works.

We note that differences in the central carbon abundance can reflect a
true physical diversity.  CO WDs of different initial masses will have
different C/O ratios in their cores.  \citet{Lesaffre2006} surveyed
the range of ignition densities expected via population synthesis of
single degenerate progenitor systems \citep[see also][]{Chen2014b}.
The spread in C/O ratio can also reflect the uncertainty associated
with the $\carbon[12](\alpha,\gamma)\oxygen[16]$ reaction rate.  For a
recent survey of how the uncertainties in this rate affect the
composition of WDs, see \citet{Fields2016}.

\subsection{Mixing Algorithm}
\label{sec:mesa-mixing}
\label{sec:mixing}

\MESA\ accounts for the mixing of species in convective regions using
a diffusive approach, in which convection gives rise to a local
diffusion coefficient $D \sim l v_\mathrm{c}$, where $l$ is the mixing
length and $v_\mathrm{c}$ the convective velocity given by mixing
length theory (MLT).  By contrast, in \PPaper\, the mixing in
convective zones is modeled as an advective process.

The mixing algorithm used in the \FUNS\ code is described in \citet{Straniero2006} as follows:
given an initial abundance $X$, the mixed abundance $X^{'}$ is
\begin{equation}
X_{j}^{'} = X_{j}+\frac{1}{M_{\mathrm{conv}}}{\sum_{k}}(X_k-X_j) f_{j,k} dm_k\,,
\end{equation}
where the $k$-summation runs over the convective zone.  The total mass
of the convective zone is $M_{\mathrm{conv}}$ and $dm_k$ is the mass
of cell $k$.  The damping factor $f$, which allows for partial mixing
in the case where the timestep is below the mixing timescale, is
\begin{equation}
  f_{j,k} = 
\begin{cases}
\frac{{\Delta}t}{{\tau}_{j,k}}  &\text{if } {\Delta}t<{\tau}_{j,k},\\
1 &\text{if } {\Delta}t\ge{\tau}_{j,k},
\end{cases}
\end{equation}
where ${\Delta}t$ is the time step and $\tau_{j,k}$ is the mixing turnover time between cells $j$ and $k$.  This quantity is evaluated as
\begin{equation}
\tau_{j,k}={\sum_{i=j}^{k}}\frac{{\Delta}r_i}{v_i},
\end{equation}
where $v_i$ is the convective velocity given by MLT.

We implement this mixing algorithm in \MESA\ using the
\texttt{other\_split\_mix} hook.  To deactivate the normal diffusive
mixing, while retaining MLT energy transport, we set
\texttt{mix\_factor = 0}.  We use \texttt{split\_mixing\_choice = -2},
which means that the mixing is applied after each timestep, in an
operator split way.

\begin{figure}[ht!]
  \includegraphics[width=\columnwidth]{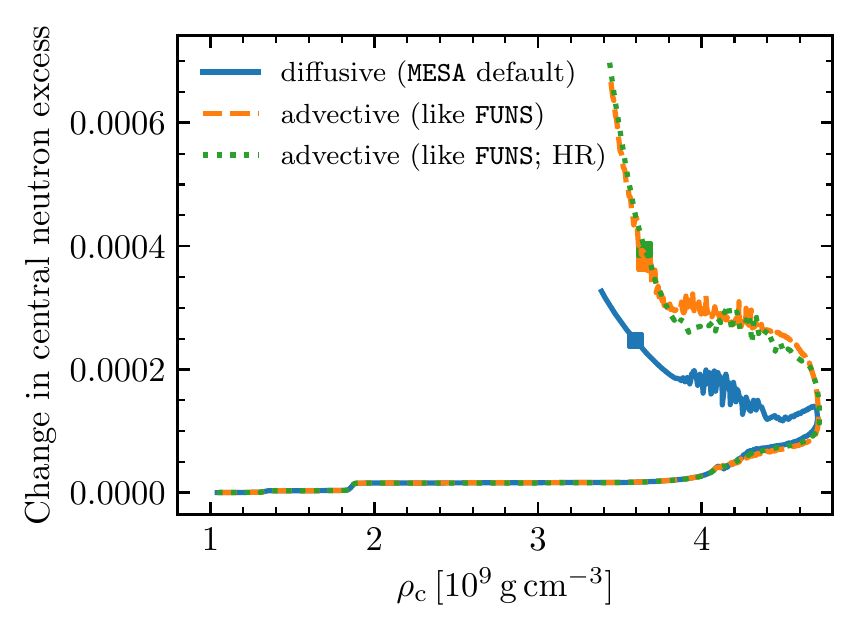}
  \caption{Change in the central neutron excess vs.~central density in
    our \MESA\ models with different mixing treatments.  Models using
    an advective mixing algorithm like that used in \PPaper\ (dashed
    line) give greater neutronization than models using the diffusive
    mixing algorithm used in \MRPaper\ (solid line).  The squares mark
    the point where $\Tc = \unit[7\times10^{8}]{K}$ and convective
    mixing is freezing out \citep{Nonaka2012}.  Both mixing treatments
    considered here are likely unreliable beyond this point.  At that
    time, the difference in neutron excess is only $\approx$ 50\%.
    For the advective mixing case, we show a high resolution run (marked HR) with
    approximately 3x greater time and space resolution; the agreement
    with the lower resolution run indicates that the models are
    numerically converged.}
\label{fig:MR16-mix-toy}
\end{figure}

In order to isolate the effects of mixing, we run models in which all
aspects are identical, except for the mixing algorithm employed.
First, we do this for the simple homogeneous models employed in the
comparison of the rates in Section~\ref{sec:simmering}.  These results
are shown in Figure~\ref{fig:MR16-mix-toy}.

\begin{figure}[ht!]
  \includegraphics[width=\columnwidth]{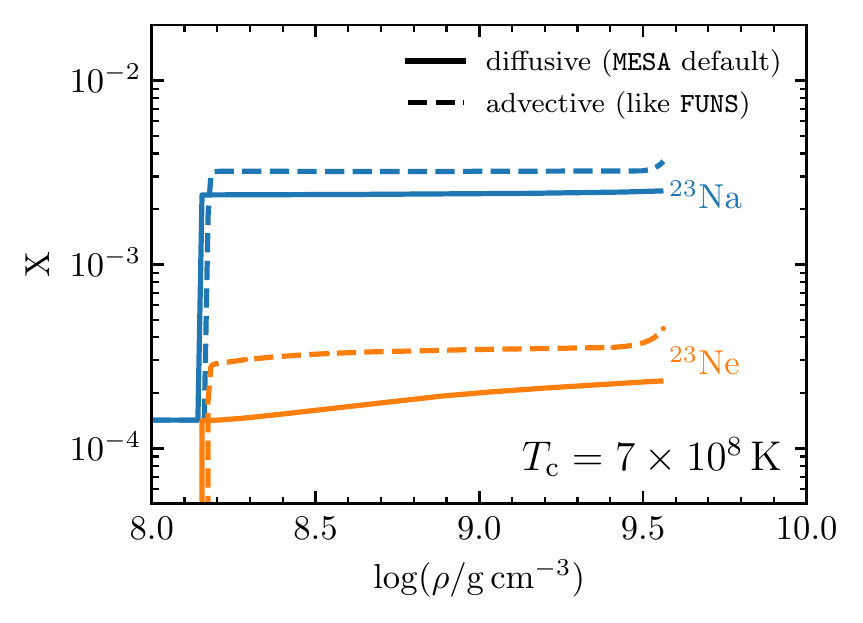}
  \caption{Composition profiles for \sodium[23] and \neon[23] in our
    \MESA\ models with different mixing treatments.  The profiles are
    shown when $\Tc = \unit[7\times10^{8}]{K}$, the point marked by
    squares in Figure~\ref{fig:MR16-mix-toy}.  Using the advective
    mixing algorithm, the mixing near the center has already begun to
    freeze out, leading to a rapid increase in the central
    neutronization beyond this point.}
\label{fig:MR16-mix-profiles}
\end{figure}

At the end of our runs, at $\Tc = \unit[8\times10^{8}]{K}$, the
difference in the neutron excess is approximately a factor of two,
with the advective mixing models favoring higher neutronizations.  We
note that 3D studies by \citet{Nonaka2012} show that mixing freezes
out around $\Tc = \unit[7\times10^{8}]{K}$.  Any treatment of mixing via MLT is
likely of questionable validity after that point.  In
Figure~\ref{fig:MR16-mix-profiles}, we show composition profiles of the
models at $\Tc = \unit[7\times10^{8}]{K}$.  This illustrates that the
two algorithms give different mixing rates and that in the advective
case, the mixing has already begun to freeze out.  This explains the
rapidly growing difference between the diffusive and advective mixing
models for $\Tc > \unit[7\times10^{8}]{K}$.  At
$\Tc = \unit[7\times10^{8}]{K}$, the difference between models with
the different mixing algorithms is $\approx$ 50\%.  This is in the
correct direction to explain the discrepancy, but is far less than the
factor of 5 difference in the reported neutronizations in \MRPaper\
and \PPaper.

\begin{figure}[ht!]
\includegraphics[width=\columnwidth]{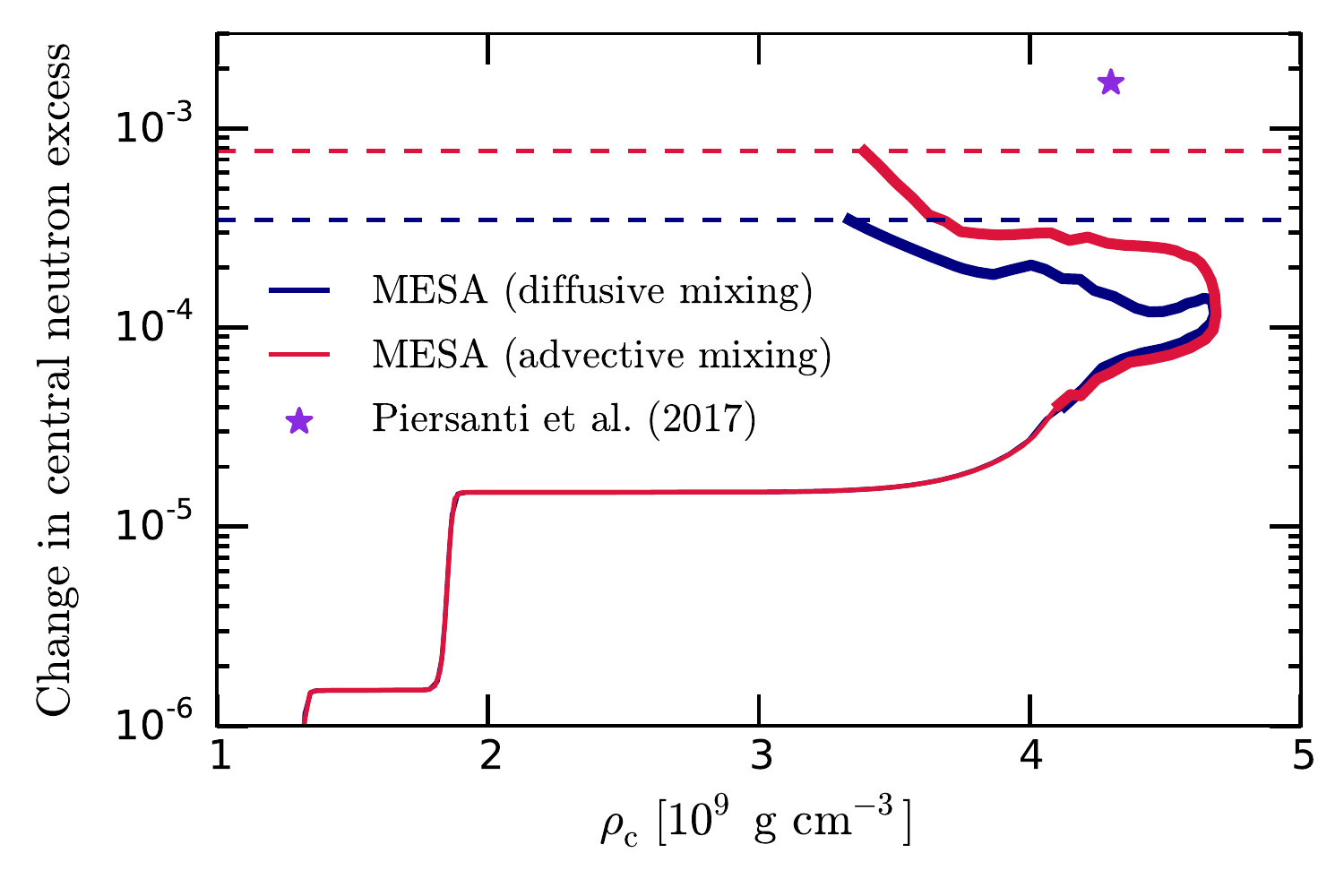}
\caption{Change in the central neutron excess vs.~central density for
  different mixing prescriptions in the selected model from MR16.  The
  convective, simmering regions are represented with thick lines.  The
  purple star depicts the ending point for the ZSUN model in \PPaper.}
\label{fig:MR16-mix}
\end{figure}

Next, we select the model from \MRPaper\ most similar to ZSUN model in
\PPaper.  This is the WD with solar metallicity, mass
$\unit[0.85]{\Msun}$, cooling age of 1 Gyr, and accretion rate of
$\unit[10^{-7}]{\Msunyr}$.  Again, we perform runs using each of the mixing algorithms.
These results are shown in
Figure~\ref{fig:MR16-mix}.  The more realistic models show a similar
increase in neutronization as in the homogenous models, demonstrating
that the homogenous models are suitable diagnostics.  While in the
correct direction, the difference is insufficient to bring us into
agreement with the final neutron excess reported by \PPaper.  This
value is shown as the star in Figure~\ref{fig:MR16-mix} and is yet a
factor of $\approx 2$ higher.  Note that the higher central density
of the \PPaper\ ending point is a consequence of the higher carbon
ignition density in those calculations (see Section~\ref{sec:simmering}).

\section{The Convective Urca Process}
\label{sec:end-simmering}

The end of simmering occurs when the central temperature reaches a
critical value $\Tc \approx \unit[8\times10^8]{K}$ \citep{Wunsch2004}.
In the absence of additional cooling or heating, the amount of carbon that burns
should be largely set by the heat capacity of the material.
\citet{Piro2008a} performed a simple calculation where they
constructed hydrostatic WD models with convective cores and examined
the thermal energy change of these models.  They then converted this
energy to an amount of carbon burned, estimating that reactions
involving \carbon[12] gave an energy $\unit[16]{MeV}$ per 6
\carbon[12] consumed.  This gave a total amount of carbon
burned $\approx \unit[2-3\times10^{-2}]{\Msun}$.
This is substantially greater than the amount of carbon burned in
\MRPaper\ and in the \MESA\ models in this paper, which is
$\approx \unit[10^{-3}]{\Msun}$.  However, it is roughly in
agreement with the \PPaper\ result,
$\unit[2.763\times10^{-3}]{\Msun}$.

\begin{figure}[ht!]
  \includegraphics[width=\columnwidth]{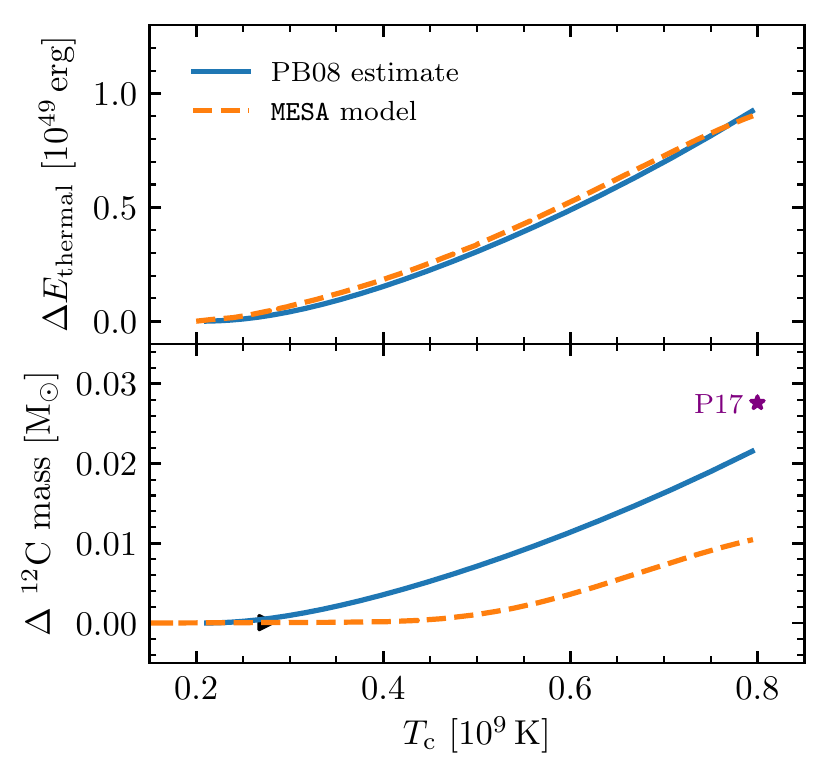}
  \caption{Change in thermal energy (top panel) and
    the amount of \carbon[12] burned (bottom panel) during the carbon simmering phase
    as a function of central temperature.  The solid lines show the
    estimates of \citet{Piro2008a} and the dashed lines show our
    fiducial \MESA\ model.  The triangle marks when the top of
    the central convective zone reaches the threshold density of
    \sodium[23]-\neon[23]. The star indicates the total amount of
    carbon burned in the ZSUN model in \PPaper.}
\label{fig:energy}
\end{figure}

The top panel in Figure~\ref{fig:energy} shows that the change in
thermal energy in our fiducial \MESA\ model and the estimate of
\citet{Piro2008a} agree.\footnote{We show their curve for a
  $\unit[1.37]{\Msun}$ WD that is initially isothermal with
  $T = \unit[2\times10^8]{K}$.}  The bottom panel shows the amount of
carbon burned.  The estimate from \citet{Piro2008a} is the thermal
energy shown in the top panel divided by their approximate specific
energy release for carbon burning of $\approx \unit[2\times10^{17}]{erg\,g^{-1}}$.
The factor of $\approx 2$ less carbon consumed in the \MESA\ model
implies the energy from carbon burning is significantly less than the change
in thermal energy.  Moreover, $\Tc$ increases from
$\unit[3\times10^8]{K}$ to $\unit[5\times10^8]{K}$ with almost no
change in the amount of carbon burned.  This indicates that another
process must be a significant source of thermal energy in our \MESA\
models.

The convective Urca process occurs when a species is advected across
its threshold density, leading to repeated electron captures and beta
decays \citep{Gamow1941, Paczynski1972b}.  For material near the
threshold density, this is a cooling process (via the neutrino
emission), but further above (below) the threshold density the
electron captures (beta decays) become exothermic \citep{Bruenn1973,
  Lesaffre2005a}.  The sideways triangle in Figure~\ref{fig:energy}
indicates the temperature at which the top of the convective zone
first reaches the threshold density of the \sodium[23]-\neon[23] Urca
pair ($\rho \approx \unit[1.8\times10^9]{\gcc}$).  Thus at
temperatures higher than this, the convective Urca process can
operate.  (We focus on \sodium[23]-\neon[23] since it is the most
abundant pair, but note that other pairs also operate once the
convection zone spans their threshold density.) The extra heating in
the \MESA\ model, beyond that supplied by carbon burning, is the
result of net heating from the convective Urca process.

In order to directly demonstrate this fact, we run \MESA\ models in
which we alter the details of the nuclear reactions beginning at the moment when a
convective core first develops.\footnote{We do not directly show the
  amount of energy from these weak reactions because it is difficult
  to extract the total energy injection from an arbitrary subset of of
  rates in \MESA.  This experiment was far simpler to do.}  Doing so leaves the
well-understood Urca-process cooling in convectively-stable regions
unchanged, as these cooling episodes occur before carbon ignition and
the onset of core convection (see Figure~\ref{fig:rates}).  If we did
not do this, the lack of cooling would shift the ignition density and
make the models more difficult to compare.

\begin{figure}
  \includegraphics[width=\columnwidth]{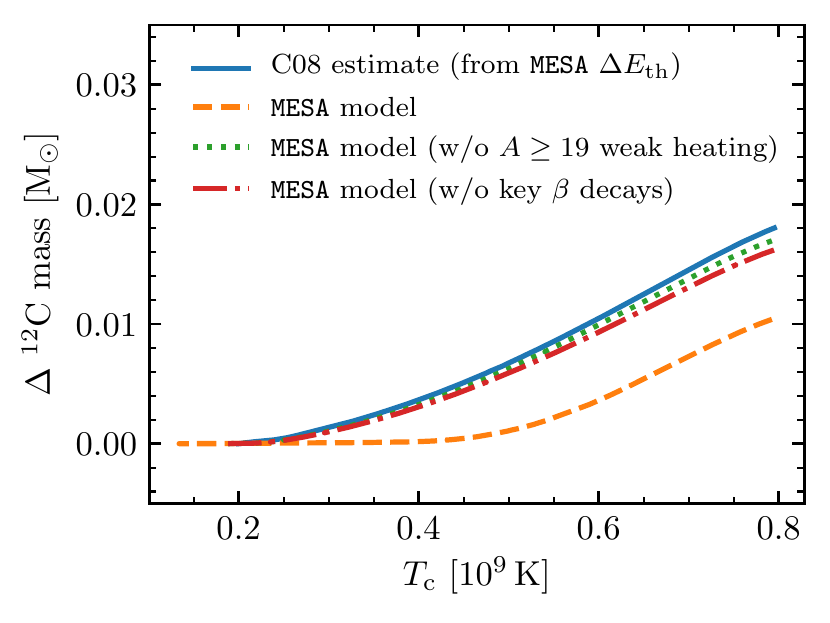}
  \caption{The amount of \carbon[12] burned during the carbon
    simmering phase as a function of central temperature.  The solid
    line shows the amount of carbon necessary to produce the change in
    thermal energy if it were the only source of heating (using the
    specific energy of carbon burning from
    \citealt{Chamulak2008}). The dashed line shows the \MESA\ model.
    The dotted and dash-dotted lines show the diagnostic models
    described in the text that remove the effects of the convective
    Urca process.}
\label{fig:energy-nwh}
\end{figure}

We show the results of these models in Figure~\ref{fig:energy-nwh}.  For
reference, the dashed line shows the amount of carbon burned in the
unmodified \MESA\ model (same as in Figure~\ref{fig:energy}).  The solid
line shows an estimate of the amount of carbon required to be burned
if all heating came from carbon burning.  This is obtained from the
change in thermal energy in the unmodified \MESA\ model and the
results of the one-zone calculations of \citet{Chamulak2008} that give
an estimate of $\unit[3.1]{MeV}$ per \carbon[12] burned.\footnote{This
  is about 15\% higher than the estimate of \citet{Piro2008a} used in
  Figure~\ref{fig:energy}; that work neglects heating from
  super-threshold electron captures, in particular
  $\nitrogen[13](e^-,\nu)\carbon[13]$ and
  $\sodium[23](e^-,\nu)\neon[23]$.}

As a first diagnostic, we run a \MESA\ model where we neglect the
energy release and neutrino losses from any weak reactions involving
isotopes with atomic number $A\ge 19$, thereby removing the thermal
effect of the convective Urca process.  In Figure~\ref{fig:energy-nwh},
the amount of carbon burned is shown as the dotted line.  In the
absence of heating from these weak reactions, carbon burning must
provide nearly all of the thermal energy necessary to raise the WD to
the final temperature.

As a second diagnostic, we run a \MESA\ model where we remove the
beta-decay reactions of \fluorine[21], \neon[23], \neon[24],
\sodium[24], \sodium[25], and \magnesium[27] from our nuclear network.
This prevents the convective Urca process from operating, but leaves
the energetic effect of the initial electron captures on these
isotopes in place.  In Figure~\ref{fig:energy-nwh}, the amount of carbon
burned is shown as the dash-dotted line.  Its similarity to the
previous case demonstrates that the heating in the unmodified model
has its origin in repeated electron captures and beta decays.

It is important to emphasize that \MESA\ models the convective Urca
process only inasmuch as it includes the appropriate weak reaction
rates.  Because \MESA\ uses a standard MLT, it does not account for the
interaction of the composition change from the weak reactions with the
convection.\footnote{\citet{Lesaffre2005a} developed a modified MLT
  for this purpose. It can be challenging to implement
  \citep{Lesaffre2004} and has not been incorporated into \MESA.}  The
decades-long debate about the effect of the convective Urca
process \citep[e.g.,][]{Paczynski1973b, Couch1974, Lazareff1975,
  Shaviv1977, Barkat1990, Mochkovitch1996, Stein2006} is ultimately a
struggle to understand this interaction.

Since the electron chemical potential increases towards the center of
the star, there must be work done as convection transports electrons
from the outer portion of the convection zone (where they are created
in beta decays) to the inner portion of the convection zone (where
they are destroyed by electron captures).  This work is not accounted
for in standard MLT, but during the run of a \MESA\ model, we can
calculate the rate at which convection is doing (unaccounted for)
work.

Similarly to \citet{Iben1978b}, we calculate the specific rate at
which this work is done as
\begin{equation}
  \epsilon_{\rm conv} = F_{\rm e} \frac{\partial \mu}{\partial m}\,,
  \label{eq:eps-conv}
\end{equation}
where $F_{\rm e}$ is the flow rate of electrons (number per time) at the given location
and $\mu$ is the electron chemical potential.  We evaluate $F_{\rm e}$
using the diffusive mass flux given by MLT in \MESA\ \citep[see Equation~14
in ][]{Paxton2011}.  This equation, plus the definition of $\Ye$ in
terms of the mass fractions $X_i$, implies that
\begin{equation}
 F_{\rm e} = \frac{\sigma}{\mpr}\frac{\partial \Ye}{\partial m}\,,
\end{equation}
where $\sigma$ is the Lagrangian diffusion coefficient associated with convection.

\begin{figure}
\includegraphics[width=\columnwidth]{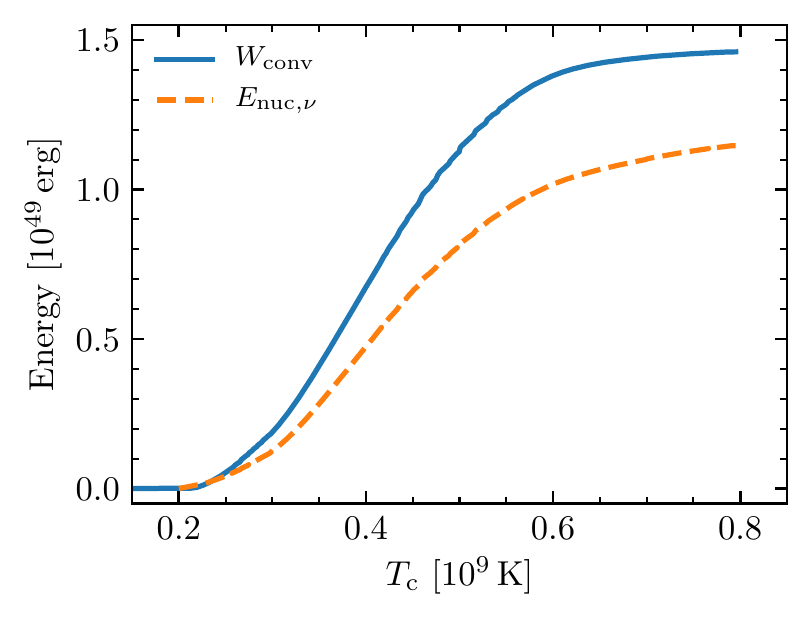}
\caption{The solid line shows the cumulative work done by convection
  as a function of central temperature.  This energy is evaluated
  using Equations~(\ref{eq:eps-conv}-\ref{eq:W-conv}), but it is not
  self-consistently included in the model.  The dashed line shows the
  cumulative nuclear neutrino losses during this phase.}
\label{fig:Urca-energetics}
\end{figure}

We define the cumulative work done by convection as
\begin{equation}
  W_{\rm conv} = \int^{t} dt' \int_0^{M_{\rm cc}} dm\, \epsilon_{\rm conv}~,
    \label{eq:W-conv}
\end{equation}
where the inner mass integral is evaluated over the central convection
zone.  The outer time integral begins at a time before a central
convection zone exists and continues until the time of interest.  We
evaluate this quantity in the \MESA\ model and show it in
Figure~\ref{fig:Urca-energetics}.

Note that this is a substantial amount of energy,
$\sim \unit[10^{49}]{erg}$, and that the work would primarily be done
when $\Tc \approx \unit[3-5\times10^{8}]{K}$.\footnote{At lower
  temperatures the convection zone has not yet encompassed the Urca
  shells; at higher temperatures the weak reactions begin to freeze
  out (see e.g., Figure~1 in \citealt{Piro2008a} or Figure~5 in
  \citealt{Chamulak2008}).}  In Figure~\ref{fig:energy}, this is
precisely when we see substantial heating unaccompanied by significant
consumption of carbon.  If the accounting were consistent, the work
that enables the convective Urca process to operate should come from
the kinetic energy of the convective motions
\citep{BisnovatyiKogan2001}, and hence it must ultimately come from
carbon burning.  That suggests, if all else were the same and the
\MESA\ MLT included this convective work term, it would be necessary
to burn an additional $\approx \unit[3\times10^{-2}]{\Msun}$ of carbon
to provide this energy.  This is roughly 3 times as much was burned in
the model, and thus would correspondingly increase the total amount of
neutronization.

The dashed line in Figure~\ref{fig:Urca-energetics} shows the cumulative
nuclear neutrino losses during the simmering phase.  Much of the
energy of sub-threshold beta decays and super-threshold electron
captures is lost to neutrinos.  The energy available for heating is
roughly the difference between the two curves, so
$\approx \unit[3\times10^{48}]{erg}$.  That corresponds to the energy
release of $\approx \unit[6\times10^{-3}]{\Msun}$ of \carbon[12],
which is consistent with the difference between the \MESA\ models with
and without the convective Urca process shown in
Figure~\ref{fig:energy-nwh}.

However, a substantial uncertainty is whether the effect of
the convective Urca process on the convective velocities serves to
limit the growth of the convective core.  In a recent study that
included calculations of the simmering phase in hybrid C/O/Ne WDs,
\citet{Denissenkov2015} show models in which they adopt a variety of
mixing assumptions including preventing the convective zone from
growing beyond the Urca shell.  In order to capture the effects of the
lingering uncertainties from the convective Urca process, future work
may want to adopt similar approaches in order to survey the range of
possible outcomes.

\section{Conclusions}
\label{sec:conclusion}

We set out to explain the differences between the results of full-star
models of the convective simmering phase of single degenerate Type Ia
supernova progenitors that were reported by
\citet{MartinezRodriguez2016} and \citet{Piersanti2017}.  We run
\MESA\ models with varying input physics assumptions.  In
Section~\ref{sec:simmering}, by incorporating the same rates used by
\PPaper, we demonstrate that the difference in reaction rates does
not make a significant contribution to the difference in neutron
excess.  In Section~\ref{sec:mesa-mixing}, we identify the details
of the mixing algorithm to be a contributor to the differences between
the results of \MRPaper\ and \PPaper.  However, this is insufficient
to explain the entire difference.

In Section~\ref{sec:end-simmering}, we demonstrate that the energy
release from electron captures and beta decays contributes
substantially to the energy budget of our \MESA\ models during
simmering.  This is a statement that in our \MESA\ models, the
convective Urca process leads to substantial net heating.  Therefore,
it is necessary to burn less carbon to deposit the necessary thermal
energy in the WD, resulting in less neutronization.

We demonstrate, from a total energy viewpoint, the importance of the
work done by convection against composition gradients.  This process
is not accounted for in the mixing length theory used in our \MESA\
models.  If it were, it would likely be necessary to burn several
times as much carbon to reach the critical temperature, resulting in
correspondingly greater neutronization.

The influence of the mixing treatment (see Section~\ref{sec:mixing})
illustrates that the thermal effects of the convective Urca process
depend on the details of how the species are mixed.  The amount of
carbon burned in the \PPaper\ models is in excess of the amount of
thermal energy needed to raise the central temperature of the WD to
$\unit[8\times10^8]{K}$, as estimated by \citet{Piro2008a} and seen in
our \MESA\ models.  One potential explanation for this would be that
the implementation details of the \FUNS\ code cause the convective Urca
process to have a net cooling effect, leading to the different
behavior seen in \PPaper.

In the end, it remains unclear what would be necessary to bring the
\MRPaper\ and \PPaper\ results into agreement.  We argue that the
neglect of convective work may lead our \MESA\ models to underestimate
the amount of carbon burned and hence the amount of neutronization.
An increase in these quantities would shift the \MESA\ models in the
direction of the \PPaper\ results.  However, it is not clear that such
a shift would bring quantitative agreement.  Additionally, a primary
finding of \PPaper\ was a strong metallicity dependence of the
neutronization, in contrast to the weak dependence found by \MRPaper.
Understanding the role of metallicity is important but is unexplored
in this study.

Substantial uncertainties associated with the convective Urca process
remain.  We emphasize that our \MESA\ models are not an entry in this
long debate.  Significant physics is missing from the \MESA\
treatment.  Instead, this work demonstrates the importance of the
development of models of the convective Urca process suitable for
integration into stellar evolution codes
\citep[e.g.,][]{Lesaffre2005a}.  While such work remains unfinished,
there may be utility in adopting simple prescriptions for the effects
of the convective Urca process \citep[e.g.,][]{Denissenkov2015}.  In
order to ensure that reported results can be compared, it will be
important to adopt approaches that give results that are not strongly
dependent on their implementation details.

\acknowledgments

We thank Stan Woosley for insightful discussions.  We thank Luciano
Piersanti and Gabriel Mart\'{i}nez-Pinedo for helpful communications
regarding the details of \citet{Piersanti2017}.  We thank Eduardo
Bravo and the anonymous referee for feedback that led to improvements
in the manuscript.  Support for this work was provided by NASA through
Hubble Fellowship grant \# HST-HF2-51382.001-A awarded by the Space
Telescope Science Institute, which is operated by the Association of
Universities for Research in Astronomy, Inc., for NASA, under contract
NAS5-26555.  H.M.-R. acknowledges support from NASA ADAP grant
NNX15AM03G S01 and a Zaccheus Daniel Predoctoral Fellowship.  This
research made extensive use of NASA's Astrophysics Data System.


\software{\MESA\ \citep{Paxton2011, Paxton2013, Paxton2015, Paxton2017},
  \texttt{Python} (available from \href{https://www.python.org}{python.org}),
  \texttt{matplotlib} \citep{matplotlib},
  \texttt{NumPy} \citep{numpy},
  \texttt{py\_mesa\_reader} \citep{pmr}
}

\bibliographystyle{aasjournal}
\bibliography{simmix}

\begin{thebibliography}{}
\expandafter\ifx\csname natexlab\endcsname\relax\def\natexlab#1{#1}\fi
\providecommand{\url}[1]{\href{#1}{#1}}

\bibitem[{{Alastuey} \& {Jancovici}(1978)}]{Alastuey1978}
{Alastuey}, A., \& {Jancovici}, B. 1978, \apj, 226, 1034

\bibitem[{{Andrews} {et~al.}(2017){Andrews}, {Weinberg}, {Sch{\"o}nrich}, \&
  {Johnson}}]{An16}
{Andrews}, B.~H., {Weinberg}, D.~H., {Sch{\"o}nrich}, R., \& {Johnson}, J.~A.
  2017, \apj, 835, 224

\bibitem[{{Arnett}(1996)}]{Arnett1996}
{Arnett}, D. 1996, {Supernovae and Nucleosynthesis: An Investigation of the
  History of Matter from the Big Bang to the Present}

\bibitem[{{Badenes} {et~al.}(2008){Badenes}, {Bravo}, \& {Hughes}}]{Ba08}
{Badenes}, C., {Bravo}, E., \& {Hughes}, J.~P. 2008, \apjl, 680, L33

\bibitem[{{Barkat} \& {Wheeler}(1990)}]{Barkat1990}
{Barkat}, Z., \& {Wheeler}, J.~C. 1990, \apj, 355, 602

\bibitem[{{Betoule} {et~al.}(2014){Betoule}, {Kessler}, {Guy}, {Mosher},
  {Hardin}, {Biswas}, {Astier}, {El-Hage}, {Konig}, {Kuhlmann}, {Marriner},
  {Pain}, {Regnault}, {Balland}, {Bassett}, {Brown}, {Campbell}, {Carlberg},
  {Cellier-Holzem}, {Cinabro}, {Conley}, {D'Andrea}, {DePoy}, {Doi}, {Ellis},
  {Fabbro}, {Filippenko}, {Foley}, {Frieman}, {Fouchez}, {Galbany}, {Goobar},
  {Gupta}, {Hill}, {Hlozek}, {Hogan}, {Hook}, {Howell}, {Jha}, {Le Guillou},
  {Leloudas}, {Lidman}, {Marshall}, {M{\"o}ller}, {Mour{\~a}o}, {Neveu},
  {Nichol}, {Olmstead}, {Palanque-Delabrouille}, {Perlmutter}, {Prieto},
  {Pritchet}, {Richmond}, {Riess}, {Ruhlmann-Kleider}, {Sako}, {Schahmaneche},
  {Schneider}, {Smith}, {Sollerman}, {Sullivan}, {Walton}, \& {Wheeler}}]{Be14}
{Betoule}, M., {Kessler}, R., {Guy}, J., {et~al.} 2014, \aap, 568, A22

\bibitem[{{Bisnovatyi-Kogan}(2001)}]{BisnovatyiKogan2001}
{Bisnovatyi-Kogan}, G.~S. 2001, \mnras, 321, 315

\bibitem[{{Bruenn}(1973)}]{Bruenn1973}
{Bruenn}, S.~W. 1973, \apjl, 183, L125

\bibitem[{{Caughlan} \& {Fowler}(1988)}]{Caughlan1988}
{Caughlan}, G.~R., \& {Fowler}, W.~A. 1988, Atomic Data and Nuclear Data
  Tables, 40, 283

\bibitem[{{Chamulak} {et~al.}(2008){Chamulak}, {Brown}, {Timmes}, \&
  {Dupczak}}]{Chamulak2008}
{Chamulak}, D.~A., {Brown}, E.~F., {Timmes}, F.~X., \& {Dupczak}, K. 2008,
  \apj, 677, 160

\bibitem[{{Chen} {et~al.}(2014){Chen}, {Han}, \& {Meng}}]{Chen2014b}
{Chen}, X., {Han}, Z., \& {Meng}, X. 2014, \mnras, 438, 3358

\bibitem[{{Chieffi} {et~al.}(1998){Chieffi}, {Limongi}, \&
  {Straniero}}]{Chieffi1998}
{Chieffi}, A., {Limongi}, M., \& {Straniero}, O. 1998, \apj, 502, 737

\bibitem[{{Couch} \& {Loumos}(1974)}]{Couch1974}
{Couch}, R.~G., \& {Loumos}, G.~L. 1974, \apj, 194, 385

\bibitem[{{Cristallo} {et~al.}(2009){Cristallo}, {Straniero}, {Gallino},
  {Piersanti}, {Dom{\'{\i}}nguez}, \& {Lederer}}]{Cristallo2009}
{Cristallo}, S., {Straniero}, O., {Gallino}, R., {et~al.} 2009, \apj, 696, 797

\bibitem[{{Denissenkov} {et~al.}(2015){Denissenkov}, {Truran}, {Herwig},
  {Jones}, {Paxton}, {Nomoto}, {Suzuki}, \& {Toki}}]{Denissenkov2015}
{Denissenkov}, P.~A., {Truran}, J.~W., {Herwig}, F., {et~al.} 2015, \mnras,
  447, 2696

\bibitem[{{Dewitt} {et~al.}(1973){Dewitt}, {Graboske}, \&
  {Cooper}}]{Dewitt1973}
{Dewitt}, H.~E., {Graboske}, H.~C., \& {Cooper}, M.~S. 1973, \apj, 181, 439

\bibitem[{{Fields} {et~al.}(2016){Fields}, {Farmer}, {Petermann}, {Iliadis}, \&
  {Timmes}}]{Fields2016}
{Fields}, C.~E., {Farmer}, R., {Petermann}, I., {Iliadis}, C., \& {Timmes},
  F.~X. 2016, \apj, 823, 46

\bibitem[{{Fuller} {et~al.}(1985){Fuller}, {Fowler}, \& {Newman}}]{Fuller1985}
{Fuller}, G.~M., {Fowler}, W.~A., \& {Newman}, M.~J. 1985, \apj, 293, 1

\bibitem[{{Gamow} \& {Schoenberg}(1941)}]{Gamow1941}
{Gamow}, G., \& {Schoenberg}, M. 1941, Physical Review, 59, 539

\bibitem[{{Graboske} {et~al.}(1973){Graboske}, {Dewitt}, {Grossman}, \&
  {Cooper}}]{Graboske1973}
{Graboske}, H.~C., {Dewitt}, H.~E., {Grossman}, A.~S., \& {Cooper}, M.~S. 1973,
  \apj, 181, 457

\bibitem[{{Hachisu} {et~al.}(1996){Hachisu}, {Kato}, \& {Nomoto}}]{Ha96}
{Hachisu}, I., {Kato}, M., \& {Nomoto}, K. 1996, \apjl, 470, L97

\bibitem[{{Han} \& {Podsiadlowski}(2004)}]{Han04}
{Han}, Z., \& {Podsiadlowski}, P. 2004, \mnras, 350, 1301

\bibitem[{Hunter(2007)}]{matplotlib}
Hunter, J.~D. 2007, Computing In Science \&amp; Engineering, 9, 90

\bibitem[{{Iben}(1978{\natexlab{a}})}]{Iben1978a}
{Iben}, Jr., I. 1978{\natexlab{a}}, \apj, 219, 213

\bibitem[{{Iben}(1978{\natexlab{b}})}]{Iben1978b}
---. 1978{\natexlab{b}}, \apj, 226, 996

\bibitem[{{Iben} \& {Tutukov}(1984)}]{Ib84}
{Iben}, Jr., I., \& {Tutukov}, A.~V. 1984, \apjs, 54, 335

\bibitem[{{Itoh} {et~al.}(1977){Itoh}, {Totsuji}, \& {Ichimaru}}]{Itoh1977}
{Itoh}, N., {Totsuji}, H., \& {Ichimaru}, S. 1977, \apj, 218, 477

\bibitem[{{Itoh} {et~al.}(1979){Itoh}, {Totsuji}, {Ichimaru}, \&
  {Dewitt}}]{Itoh1979}
{Itoh}, N., {Totsuji}, H., {Ichimaru}, S., \& {Dewitt}, H.~E. 1979, \apj, 234,
  1079

\bibitem[{{Kobayashi} {et~al.}(2006){Kobayashi}, {Umeda}, {Nomoto}, {Tominaga},
  \& {Ohkubo}}]{Ko06}
{Kobayashi}, C., {Umeda}, H., {Nomoto}, K., {Tominaga}, N., \& {Ohkubo}, T.
  2006, \apj, 653, 1145

\bibitem[{{Kushnir} {et~al.}(2013){Kushnir}, {Katz}, {Dong}, {Livne}, \&
  {Fern{\'a}ndez}}]{Ku13}
{Kushnir}, D., {Katz}, B., {Dong}, S., {Livne}, E., \& {Fern{\'a}ndez}, R.
  2013, \apjl, 778, L37

\bibitem[{{Lazareff}(1975)}]{Lazareff1975}
{Lazareff}, B. 1975, \aap, 45, 141

\bibitem[{{Lesaffre} {et~al.}(2006){Lesaffre}, {Han}, {Tout}, {Podsiadlowski},
  \& {Martin}}]{Lesaffre2006}
{Lesaffre}, P., {Han}, Z., {Tout}, C.~A., {Podsiadlowski}, P., \& {Martin},
  R.~G. 2006, \mnras, 368, 187

\bibitem[{{Lesaffre} {et~al.}(2005){Lesaffre}, {Podsiadlowski}, \&
  {Tout}}]{Lesaffre2005a}
{Lesaffre}, P., {Podsiadlowski}, P., \& {Tout}, C.~A. 2005, \mnras, 356, 131

\bibitem[{{Lesaffre} {et~al.}(2004){Lesaffre}, {Tout}, {Stancliffe}, \&
  {Podsiadlowski}}]{Lesaffre2004}
{Lesaffre}, P., {Tout}, C.~A., {Stancliffe}, R.~J., \& {Podsiadlowski}, P.
  2004, \memsai, 75, 660

\bibitem[{{Malone} {et~al.}(2014){Malone}, {Nonaka}, {Woosley}, {Almgren},
  {Bell}, {Dong}, \& {Zingale}}]{Malone2014a}
{Malone}, C.~M., {Nonaka}, A., {Woosley}, S.~E., {et~al.} 2014, \apj, 782, 11

\bibitem[{{Mart{\'{\i}}nez-Rodr{\'{\i}}guez}
  {et~al.}(2016){Mart{\'{\i}}nez-Rodr{\'{\i}}guez}, {Piro}, {Schwab}, \&
  {Badenes}}]{MartinezRodriguez2016}
{Mart{\'{\i}}nez-Rodr{\'{\i}}guez}, H., {Piro}, A.~L., {Schwab}, J., \&
  {Badenes}, C. 2016, \apj, 825, 57

\bibitem[{{Mart{\'{\i}}nez-Rodr{\'{\i}}guez}
  {et~al.}(2017){Mart{\'{\i}}nez-Rodr{\'{\i}}guez}, {Badenes}, {Yamaguchi},
  {Bravo}, {Timmes}, {Miles}, {Townsley}, {Piro}, {Mori}, {Andrews}, \&
  {Park}}]{MartinezRodriguez2017}
{Mart{\'{\i}}nez-Rodr{\'{\i}}guez}, H., {Badenes}, C., {Yamaguchi}, H.,
  {et~al.} 2017, \apj, 843, 35

\bibitem[{{Mochkovitch}(1996)}]{Mochkovitch1996}
{Mochkovitch}, R. 1996, \aap, 311, 152

\bibitem[{{Nomoto} {et~al.}(1984){Nomoto}, {Thielemann}, \& {Yokoi}}]{No84}
{Nomoto}, K., {Thielemann}, F.-K., \& {Yokoi}, K. 1984, \apj, 286, 644

\bibitem[{{Nonaka} {et~al.}(2012){Nonaka}, {Aspden}, {Zingale}, {Almgren},
  {Bell}, \& {Woosley}}]{Nonaka2012}
{Nonaka}, A., {Aspden}, A.~J., {Zingale}, M., {et~al.} 2012, \apj, 745, 73

\bibitem[{{Paczy{\'n}ski}(1972)}]{Paczynski1972b}
{Paczy{\'n}ski}, B. 1972, \aplett, 11, 53

\bibitem[{{Paczy{\'n}ski}(1973{\natexlab{a}})}]{Paczynski1973a}
---. 1973{\natexlab{a}}, \actaa, 23, 1

\bibitem[{{Paczy{\'n}ski}(1973{\natexlab{b}})}]{Paczynski1973b}
---. 1973{\natexlab{b}}, \aplett, 15, 147

\bibitem[{{Park} {et~al.}(2013){Park}, {Badenes}, {Mori}, {Kaida}, {Bravo},
  {Schenck}, {Eriksen}, {Hughes}, {Slane}, {Burrows}, \& {Lee}}]{Park13}
{Park}, S., {Badenes}, C., {Mori}, K., {et~al.} 2013, \apjl, 767, L10

\bibitem[{{Paxton} {et~al.}(2011){Paxton}, {Bildsten}, {Dotter}, {Herwig},
  {Lesaffre}, \& {Timmes}}]{Paxton2011}
{Paxton}, B., {Bildsten}, L., {Dotter}, A., {et~al.} 2011, \apjs, 192, 3

\bibitem[{{Paxton} {et~al.}(2013){Paxton}, {Cantiello}, {Arras}, {Bildsten},
  {Brown}, {Dotter}, {Mankovich}, {Montgomery}, {Stello}, {Timmes}, \&
  {Townsend}}]{Paxton2013}
{Paxton}, B., {Cantiello}, M., {Arras}, P., {et~al.} 2013, \apjs, 208, 4

\bibitem[{{Paxton} {et~al.}(2015){Paxton}, {Marchant}, {Schwab}, {Bauer},
  {Bildsten}, {Cantiello}, {Dessart}, {Farmer}, {Hu}, {Langer}, {Townsend},
  {Townsley}, \& {Timmes}}]{Paxton2015}
{Paxton}, B., {Marchant}, P., {Schwab}, J., {et~al.} 2015, \apjs, 220, 15

\bibitem[{{Paxton} {et~al.}(2016){Paxton}, {Marchant}, {Schwab}, {Bauer},
  {Bildsten}, {Cantiello}, {Dessart}, {Farmer}, {Hu}, {Langer}, {Townsend},
  {Townsley}, \& {Timmes}}]{Paxton2016}
---. 2016, \apjs, 223, 18

\bibitem[{{Paxton} {et~al.}(2017){Paxton}, {Schwab}, {Bauer}, {Bildsten},
  {Blinnikov}, {Duffell}, {Farmer}, {Goldberg}, {Marchant}, {Sorokina},
  {Thoul}, {Townsend}, \& {Timmes}}]{Paxton2017}
{Paxton}, B., {Schwab}, J., {Bauer}, E.~B., {et~al.} 2017, ArXiv e-prints,
  arXiv:1710.08424

\bibitem[{{Perlmutter} {et~al.}(1999){Perlmutter}, {Aldering}, {Goldhaber},
  {Knop}, {Nugent}, {Castro}, {Deustua}, {Fabbro}, {Goobar}, {Groom}, {Hook},
  {Kim}, {Kim}, {Lee}, {Nunes}, {Pain}, {Pennypacker}, {Quimby}, {Lidman},
  {Ellis}, {Irwin}, {McMahon}, {Ruiz-Lapuente}, {Walton}, {Schaefer}, {Boyle},
  {Filippenko}, {Matheson}, {Fruchter}, {Panagia}, {Newberg}, {Couch}, \&
  {Project}}]{Pe99}
{Perlmutter}, S., {Aldering}, G., {Goldhaber}, G., {et~al.} 1999, \apj, 517,
  565

\bibitem[{{Piersanti} {et~al.}(2017){Piersanti}, {Bravo}, {Cristallo},
  {Dom{\'{\i}}nguez}, {Straniero}, {Tornamb{\'e}}, \&
  {Mart{\'{\i}}nez-Pinedo}}]{Piersanti2017}
{Piersanti}, L., {Bravo}, E., {Cristallo}, S., {et~al.} 2017, \apjl, 836, L9

\bibitem[{{Piro} \& {Bildsten}(2008)}]{Piro2008a}
{Piro}, A.~L., \& {Bildsten}, L. 2008, \apj, 673, 1009

\bibitem[{{Piro} \& {Chang}(2008)}]{Piro2008b}
{Piro}, A.~L., \& {Chang}, P. 2008, \apj, 678, 1158

\bibitem[{{Rest} {et~al.}(2014){Rest}, {Scolnic}, {Foley}, {Huber}, {Chornock},
  {Narayan}, {Tonry}, {Berger}, {Soderberg}, {Stubbs}, {Riess}, {Kirshner},
  {Smartt}, {Schlafly}, {Rodney}, {Botticella}, {Brout}, {Challis}, {Czekala},
  {Drout}, {Hudson}, {Kotak}, {Leibler}, {Lunnan}, {Marion}, {McCrum},
  {Milisavljevic}, {Pastorello}, {Sanders}, {Smith}, {Stafford}, {Thilker},
  {Valenti}, {Wood-Vasey}, {Zheng}, {Burgett}, {Chambers}, {Denneau}, {Draper},
  {Flewelling}, {Hodapp}, {Kaiser}, {Kudritzki}, {Magnier}, {Metcalfe},
  {Price}, {Sweeney}, {Wainscoat}, \& {Waters}}]{Re14}
{Rest}, A., {Scolnic}, D., {Foley}, R.~J., {et~al.} 2014, \apj, 795, 44

\bibitem[{{Riess} {et~al.}(1998){Riess}, {Filippenko}, {Challis},
  {Clocchiatti}, {Diercks}, {Garnavich}, {Gilliland}, {Hogan}, {Jha},
  {Kirshner}, {Leibundgut}, {Phillips}, {Reiss}, {Schmidt}, {Schommer},
  {Smith}, {Spyromilio}, {Stubbs}, {Suntzeff}, \& {Tonry}}]{Ri98}
{Riess}, A.~G., {Filippenko}, A.~V., {Challis}, P., {et~al.} 1998, \aj, 116,
  1009

\bibitem[{{Shaviv} \& {Regev}(1977)}]{Shaviv1977}
{Shaviv}, G., \& {Regev}, O. 1977, \aap, 54, 581

\bibitem[{{Shen} {et~al.}(2017){Shen}, {Kasen}, {Miles}, \&
  {Townsley}}]{Shen2017b}
{Shen}, K.~J., {Kasen}, D., {Miles}, B.~J., \& {Townsley}, D.~M. 2017, ArXiv
  e-prints, arXiv:1706.01898

\bibitem[{{Sim} {et~al.}(2010){Sim}, {R{\"o}pke}, {Hillebrandt}, {Kromer},
  {Pakmor}, {Fink}, {Ruiter}, \& {Seitenzahl}}]{Si10}
{Sim}, S.~A., {R{\"o}pke}, F.~K., {Hillebrandt}, W., {et~al.} 2010, \apjl, 714,
  L52

\bibitem[{{Stein} \& {Wheeler}(2006)}]{Stein2006}
{Stein}, J., \& {Wheeler}, J.~C. 2006, \apj, 643, 1190

\bibitem[{{Straniero} {et~al.}(2006){Straniero}, {Gallino}, \&
  {Cristallo}}]{Straniero2006}
{Straniero}, O., {Gallino}, R., \& {Cristallo}, S. 2006, Nuclear Physics A,
  777, 311

\bibitem[{{Suzuki} {et~al.}(2016){Suzuki}, {Toki}, \& {Nomoto}}]{Suzuki2016}
{Suzuki}, T., {Toki}, H., \& {Nomoto}, K. 2016, \apj, 817, 163

\bibitem[{{Toki} {et~al.}(2013){Toki}, {Suzuki}, {Nomoto}, {Jones}, \&
  {Hirschi}}]{Toki2013}
{Toki}, H., {Suzuki}, T., {Nomoto}, K., {Jones}, S., \& {Hirschi}, R. 2013,
  \prc, 88, 015806

\bibitem[{{Tsuruta} \& {Cameron}(1970)}]{Tsuruta1970}
{Tsuruta}, S., \& {Cameron}, A.~G.~W. 1970, \apss, 7, 374

\bibitem[{van~der Walt {et~al.}(2011)van~der Walt, Colbert, \&
  Varoquaux}]{numpy}
van~der Walt, S., Colbert, S.~C., \& Varoquaux, G. 2011, Computing in Science
  Engineering, 13, 22

\bibitem[{Wolf \& Schwab(2017)}]{pmr}
Wolf, B., \& Schwab, J. 2017, wmwolf/py\_mesa\_reader: Interact with MESA
  Output, , , doi:10.5281/zenodo.826958.
\newblock \url{https://doi.org/10.5281/zenodo.826958}

\bibitem[{{Woosley} {et~al.}(2004){Woosley}, {Wunsch}, \&
  {Kuhlen}}]{Woosley2004a}
{Woosley}, S.~E., {Wunsch}, S., \& {Kuhlen}, M. 2004, \apj, 607, 921

\bibitem[{{Wunsch} \& {Woosley}(2004)}]{Wunsch2004}
{Wunsch}, S., \& {Woosley}, S.~E. 2004, \apj, 616, 1102

\bibitem[{{Yakovlev} {et~al.}(2006){Yakovlev}, {Gasques}, {Afanasjev}, {Beard},
  \& {Wiescher}}]{Yakovlev2006}
{Yakovlev}, D.~G., {Gasques}, L.~R., {Afanasjev}, A.~V., {Beard}, M., \&
  {Wiescher}, M. 2006, \prc, 74, 035803

\bibitem[{{Yamaguchi} {et~al.}(2015){Yamaguchi}, {Badenes}, {Foster}, {Bravo},
  {Williams}, {Maeda}, {Nobukawa}, {Eriksen}, {Brickhouse}, {Petre}, \&
  {Koyama}}]{Ya15}
{Yamaguchi}, H., {Badenes}, C., {Foster}, A.~R., {et~al.} 2015, \apjl, 801, L31

\bibitem[{{Zegers} {et~al.}(2008){Zegers}, {Brown}, {Akimune}, {Austin}, {van
  den Berg}, {Brown}, {Chamulak}, {Fujita}, {Fujiwara}, {Gal{\`e}s}, {Harakeh},
  {Hashimoto}, {Hayami}, {Hitt}, {Itoh}, {Kawabata}, {Kawase}, {Kinoshita},
  {Nakanishi}, {Nakayama}, {Okumura}, {Shimbara}, {Uchida}, {Ueno}, {Yamagata},
  \& {Yosoi}}]{Zegers2008}
{Zegers}, R.~G.~T., {Brown}, E.~F., {Akimune}, H., {et~al.} 2008, \prc, 77,
  024307

\end{thebibliography}

\end{document}